\documentclass[preprint,showpacs,preprintnumbers,amsmath,amssymb]{revtex4}

\begin{document}
\title{ Discrete Quantum Spectrum of  Observable Correlations from Inflation}
\author{Craig J. Hogan}
\address{Astronomy and Physics Departments, 
University of Washington,
Seattle, Washington 98195-1580}
\begin{abstract}

The decoherence of quantum fluctuations into classical  perturbations during inflation is discussed. A simple quantum mechanical argument, using a spatial particle wavefunction  rather than a field description, shows that observable correlations from inflation must have a discrete spectrum, since they originate and  freeze into the metric within a compact region.  The number of discrete modes is estimated using a  holographic bound  on the number of degrees of freedom. The discreteness may be detectable  in some models; for example, if there is a  fundamental universal frequency spectrum, the inflationary gravitational wave background may be resolvable into discrete emission lines.
\end{abstract}
\pacs{98.80.-k}
\maketitle
\section{introduction}

Various approaches to quantum gravity--- black hole states in string theory, holographic entropy bounds, dualities of gravity states in anti-de Sitter and de Sitter spaces with systems of fields living in projective spaces in smaller dimensionalities, loop quantum gravity--- suggest that nature is fundamentally discrete, and that quantum systems of fields with gravity contain  a finite amount of information\cite{Maldacena:1997re,'tHooft:1999bw,'tHooft:1985re,Susskind:1993if,susskind95,bousso02,Alishahiha:2004md,Rovelli:1997yv}. For example,  black hole states and evaporation radiation are conjectured to have a discrete spectrum\cite{Bekenstein:1997bt}. It is  generally thought that signs of the effect are not detectable in practice  in the present-day universe.

On the other hand, large scale cosmic perturbations have a quantum structure frozen into the spacetime metric under extreme conditions, in a very compact volume, when the information content was much more limited than it is today.  Although it has not been clear how this information limit would manifest itself observationally, a discrete spectrum in spatial structure may in principle be detectable.\cite{Hogan:2003mq,Hogan:2004vi} The standard inflation theory\cite{lythriotto,liddlelyth,Starobinsky:1979ty,Hawking:1982cz,Guth:1982ec,Bardeen:1983qw,Starobinsky:1982ee,Halliwell:1985eu,guthpi,Grishchuk:1989ss,Grishchuk:1990bj,Albrecht:1994kf,Lesgourgues:1997jc,Polarski:1996jg,kiefer,Kiefer:1999sj}, which treats the background spacetime classically,  predicts a continuous spectrum, with an infinite information content, so new physics must be added to analyze the phenomenology of this effect.

A new approach is adopted here to analyze the observable structural  relics of limited frozen information as spatial correlations of  quantum mechanical particles, rather than independently evolved field modes. This view allows application of a simple and general property of quanta, due originally to Ehrenfest: 
 confined or spatially  bounded states always  have a discrete spectrum. Inflationary correlations arise and freeze in within a small causally connected volume and therefore have a discrete spectrum.

This result is used to argue that localization in space 
of observable metric perturbations at the time perturbations are frozen in implies a discrete $\vec k$ spectrum  that survives to late times. Quantitative estimates based the covariant entropy bound  indicate that  the effect is accessible in principle, though it  is likely to be detectable in practice only in some models.   In particular, if  there is a  fundamental universal frequency spectrum, it may be seen  by experiments now being contemplated.

\section{quantum measurement theory of inflationary fields}

A beautiful result  of inflation theory is the natural conversion of quantum fluctuations  into classical perturbations.
 An initial vacuum state  of a field is defined in the usual way using annihilation operators on field modes.  At the initial stage the expansion of the spacetime is negligible, so the quantum mechanics of each mode corresponds in the usual way to a simple harmonic oscillator. The initial vacuum or ground state is  a definite eigenstate with  zero mean particle occupation number, but nonzero fluctuations in field amplitude.
 The physical size of a comoving mode  expands with the spacetime, and its frequency decreases. Eventually its wavelength becomes larger than the apparent horizon and its oscillation frequency much less than the expansion rate. The eigenstates for the  mode are then no longer those  of definite particle occupation number, but of definite field amplitude.  An observation at late times necessarily finds each mode of the field in one of these eigenstates, corresponding to a definite classical field configuration, which in turn  unambiguously maps onto the classical spatial perturbation of the spacetime metric.
 
The standard theory allows a clean calculation of the probability distribution of mode amplitudes\cite{Guth:1982ec}. 
No subtle quantum measurement theory is required to compute statistical observables, such as the mean power spectrum, to compare with measured statistical properties  of the universe. However, observable such as the  specific,  final locations of actual galaxies, does depend on the decoherence or effective measurement process.

The apparent  (but illusory) decoherence of the initial quantum state does not require any actual observers to be present at the time the modes freeze into classical perturbations, nor does any special collapse process need to be invoked. Since eigenstates correspond  to  definite field amplitude,  the wavefunction of the whole system branches in such a way that the field amplitudes simply appear classical, like the final states of  Schr\"odinger's cat.

As in the case of the cat, the wavefunction of the system does not really collapse. It maintains a superposition of live cats and dead cats, each correlated with a corresponding happy or sad owner sharing the same branch of the wavefunction. 
The full wavefunction of the inflationary field  similarly maintains a quantum superposition 
of different field amplitudes and phases for each mode, although any observer (or any late-time metric realization) only sees one of them. This view of decoherence is analyzed via  correlations of coarse-grained histories\cite{Gell-Mann:1995cu,Hartle:2002nq,Bosse:2005un}. The main difference between the cat and the field is that the cat requires an apparatus to connect a quantum decay to a classical cat state, whereas the field mode  changes from  quantum to quasi-classical behavior on its own, due to the expansion of the spacetime. 

There is therefore still an aspect of the physics of the system that is not yet being captured by the standard inflation+field formalism: spatial localization of frozen field correlations. When an inflationary field mode freezes out, in the standard formalism its state freezes out uniformly across all of space. But physically, correlations in the metric created by the mode can only be    frozen into the metric (and ``remembered'' for later observations) over localized regions, corresponding to the comoving size of the apparent horizon at the time of freezing. 

Put another way, field theory predicts the creation of particles, but its choice of  exactly where in space they should go depends on  the phases of field modes that stretch to infinity. That choice (which corresponds to choosing where hot and cold spots go in the microwave sky, or where galaxies end up collapsing) is made very early, at the time of mode freezing, when there is no physical way for the frozen-in phases of field modes to be correlated over larger distances than a few wavelengths. 
This aspect of the quantum behavior can be more easily described using a spatial particle wavefunction, instead of a quantized field. 

\section{spatially bounded correlation implies discrete spectrum}
Consider a particle wavefunction $\Psi(x)$ on one infinite spatial dimension: the quantum-mechanical amplitude for a particle to be at point $x$. At each point,  $|\Psi(x)|^2$ is proportional to the probability density to find a particle at that point. In general, the wavefunction may be decomposed into a  spectrum:
\begin{equation}
\Psi(x) = \int dk e^{ikx}\Psi_k,
\end{equation}

Statistical structure can be  described by a two-point  wavefunction, $\Psi(x)\Psi^*(x+y)$.  In a condensate where many particles share  the same wavefunction, $\Psi(x)\Psi^*(x+y)$ is  the quantum-mechanical amplitude to find a particle at $x$ and also one at $x+y$. Define a correlation wavefunction
\begin{equation}
f(y)= \langle \Psi(x)\Psi^*(x+y)\rangle_x,
\end{equation}
 where brackets indicate a spatial average over $x$. 
  Physically, $|f(y)|^2$ is another observable,  proportional to the probability of finding a particle at position $y$ from a given particle.  In the inflationary context described below, we will assume that  it is proportional at late times to  the classical correlation function for  metric perturbations. Therefore, apart from transfer effects at late times, $|f(y)|^2$ corresponds to the correlations observed on any branch of the wavefunction, and its transform corresponds to the observed power spectrum.

The spectral decomposition of $f(y)$  is then: 
\begin{equation}
f(y)
 =\int\int dk dk' \langle\exp [ix(k-k'){-ik'y}]\rangle_x \Psi_{k}\Psi^*_{k'},
=\int dk e^{-iky} \Psi_k\Psi_k^*
\end{equation}
 
 Suppose that $f(y)$ has compact support--- that it vanishes outside of some interval of width $Y_0$.  Then  the coefficient  $\Psi_k\Psi_k^*$ vanishes except for $k$ having integer multiples of $k_0=(\pi Y_0)^{-1}$:  its spectrum is discrete.
 Conversely,  a continuous spectrum  implies infinite support in $y$. 

Therefore, in a system that yields spatially bounded observable correlations between particles, those correlations have a  discrete spatial spectrum. Without loss of generality,
 \begin{equation}\label{eqn:spectrum}
f(y)= \langle\Psi(x)\Psi^*(x+y)\rangle
 =\sum_{n=1}^N e^{-ik_ny}   f_{k_n}.
\end{equation}
 This derivation is formally only slightly different from the standard result in quantum mechanics that bound states always have a discrete energy spectrum.\cite{LL}
 In general the  spectrum  in energy and momentum, and the spatial structure of the eigenmodes,  depend  on the Hamiltonian of the quantum system. 
As discussed below, 
there are also reasons to expect that $N$ is finite in the inflationary system.
  
\section{Application to inflation}

Consider the case where $\Psi(x)$ is the 
 wavefunction of the inflaton or graviton during inflation.
 The single dimension can be chosen to be any   direction on the standard spatial hypersurfaces used to describe field modes during inflation. The coordinate $x$ can be taken to be a comoving coordinate on one of these hypersurfaces, and $k$ a comoving wavenumber.
 
In standard inflation theory, the comoving spectrum of observed  field modes  is determined at the epoch when they freeze out, at about the same time they cross the horizon.  After this time, their spatial structure behaves in an entirely classical way, and they can be described as part of the classical background metric.  The  freezing fixes the contribution of modes of a given scale to the spectrum of $f(y)$.
 
The quantum-mechanical correlations between particles  generated  
 during this process, and described by $f(y)$, are limited by causality. The physical  diameter of the apparent horizon  during inflation has a proper length $2/H$, where $H$ is the expansion rate during inflation.  Since modes freeze into the metric when they are of the order of this scale, all nonzero, observable correlations  in the final classical metric are introduced within a finite region of about this size. According to the above argument, they therefore have a discrete spectrum in $k$. 
 The freezing of modes at a certain epoch on inflationary hypersurfaces implies that a discrete spectrum is fixed at that time.

Briefly: at the time when modes of any given comoving scale freeze into their observed classical spectrum, correlations due to these modes vanish on scales much larger than the comoving size of the horizon at that time; therefore, they have a discrete spectrum.

(Note that one would be  free to choose a different, highly curved spacelike hypersurface with a much larger proper size. However, this choice   mixes states of field modes freezing out at different times in the standard coordinates, and the $k$'s do not map onto states  of field modes in the standard coordinates that eventually map onto spatial modes of observable metric perturbations at late times.)
 
The role of the spacetime metric in particle localization can be thought of metaphorically in terms  of  the uncertainty principle.
Long before a mode  freezes out, in the limit where the expansion can be ignored, the field is in a vacuum state. The emergence of the discrete spectrum goes hand in hand with the creation of `particles' from the vacuum via `interaction' with the metric curvature. Structure freezes into the metric  at a field fluctuation amplitude corresponding to creation of  about one quantum of action per horizon volume. The spacetime metric ``observes" the field, ``collapsing" the wavefunction (and the metric) into a definite spatial configuration, and in so doing, kicks the field into an excited state with nonzero occupation number, creating localized particles from the vacuum. The creation of quanta at a definite spacetime location collapses the phases of field modes on a comparable scale on a particular branch of the wavefunction.

 Thus,  the highly curved inflationary spacetime freezes out the modes and   localizes the perturbations.  
 The discreteness  can be viewed as an effect of the quanta of the fields interacting with those of the spacetime as the perturbation becomes part of the classical background. With a discrete spectrum, it is possible to collapse the wavefunction with locally obtainable information.
 
 The connection of discreteness and localization appears to be a new effect, not incorporated into the standard inflation theory.   It does not affect the key prediction of mean  broad-band power spectrum from standard  inflation theory, still an excellent approximation for average fluctuation power.

Standard slow-roll inflation is approximately scale-free and isotropic. These symmetries also apply to $f(y)$ and its transform $f_k$.  From these symmetries we conclude that along any axis through the origin in $k$ space, $f_k$ has discrete spectral ``lines'' equally spaced in $\log k$. However, to estimate their spacing, or  properties in other dimensions, we would need to introduce more physics.

\section{Holographic estimates of observable discreteness}

 It appears that spectral discreteness of the frozen correlations--- implying a countable number of frozen inflationary spatial modes---  is induced by spacetime causal structure.    At the same time, exact ``holographic'' correspondences have been conjectured, connecting a spacetime's causal structure with a limit on the total number of quantum degrees of freedom.\cite{Maldacena:1997re,'tHooft:1999bw,'tHooft:1985re,Susskind:1993if,susskind95,bousso02,Alishahiha:2004md}  Based on the assumption  that these effects are related,     holographic arguments   provide a quantitative  estimate of properties of the discrete spatial spectrum from inflation. 
 
In field theory, each spatial mode  represents an approximately independent degree of freedom, and their sum completely describes the field.  It is natural then to identify the number of (frozen) degrees of freedom at any time, the number of discrete classical modes,  with  the number of quantum degrees of freedom, as limited by holography.

More precisely, the covariant entropy bound  applied to each quasi-de Sitter static patch of an inflationary universe gives a  maximum entropy $\pi m_P^2/H^2$ per 3-volume $4\pi/3H^2$. Conservatively, this can be used to derive a bound on 
information density ${\cal I} $  for approximately isotropic, scale-invariant inflationary modes that freeze  in as classical perturbations at late times\cite{Hogan:2004vi}: 
 \begin{equation}\label{eqn:ibound}
{\cal I}    =   (9/16\pi) {\cal F} m_P^2 H^{-2},
\end{equation}
in dimensionless phase space (spatial times wavenumber volume),
where  ${\cal F}<1$ denotes the fraction of the covariant bound on entropy
 carried by the information in the frozen field modes with $k<k'$ at a time when ${k'}<H$. The factor $m_P^2/ H^{2}$ exceeds $10^6$ but could be much greater. 
 
For general models saturating this conservative bound, discreteness may be difficult or impossible to observe.  For example, consider an experiment to directly detect gravitational waves from inflationary gravitons in a broad band at frequency $\omega$. Suppose the experiment runs for a time $\tau$ and attains frequency resolution $\delta\omega\approx \tau^{-1}$.
The bound (\ref{eqn:ibound})
permits  ${\cal I} (\omega\tau)^3$ different observable modes in this experimental phase space. Assuming they all have different frequencies, the spectrum is resolved into discrete frequencies only  if $\omega/\delta\omega>{\cal I}  (\omega\tau)^3$, requiring ${\cal I}< (\omega\tau)^{-2}$, which is always less than unity.  
More generally, even in the case of surveys where all spatial directions can be distinguished observationally, spectral discreteness can be resolved only if
 ${\cal F}$ is small enough that ${\cal I} $ is of the order of unity or less.

However, discreteness is much easier to detect in some models.  For example, suppose that  there is a fundamental, universal frequency spectrum(see e.g., \cite{Seahra:2004fg}), with a mode density  in $k$ space that still obeys the holographic bound in each de Sitter volume during inflation.  This possibility is at least suggested by holographic dual models of de Sitter space\cite{Alishahiha:2004md}. The lowest frequency states including gravity in a 3+1 de Sitter horizon volume correspond to the  highest frequency states of a conformal field theory living  in the cross-sectional 2+1-dimensional de Sitter horizon volume. The freezing of degrees of freedom as modes expand beyond the horizon  in the dS3+1 corresponds to structures shrinking below the Planck scale in the dS2+1; after this, they no longer participate as quantum degrees of freedom but are part of the frozen classical background structure.  The dS2+1 maximum frequency is presumably a fundamental universal scale, perhaps corresponding to a fundamental lowest frequency (and frequency spacing) for states in the dS3+1 space where inflationary modes live. 

 In this situation, for each $H$ there is some fixed number of mode frequencies $N$  in Eq. (\ref{eqn:spectrum}),  rather than an information density in $k$ space that increases with $x$ volume. As part of the freezing process for modes, $H$ itself changes in discrete steps.\cite{Zizzi:1999sx}  Holographic entropy bounds then suggest a scale-invariant spectrum with $N< \pi m_P^2/H^2$ modes per octave of frequency, which might be as few as $10^6$. 
 A variety of tests might reveal discreteness at this level\cite{Hogan:2003mq}.
  
 In particular, a gravitational wave
experiment  resolves $N\approx \omega/\delta\omega$, which is of the order of $10^8$ for a 1 Hz spaceborne experiment such as the Big Bang Observer concept mission\cite{Hogan:2003mq}.    Rather than the continuous Gaussian noise predicted in standard inflation, the background appears in discrete, well separated spectral lines occupying a small fraction of the spectrum. The concentration of power in narrow bands actually makes detection easier
since the signal-to-noise ratio in the lines is larger than the continuum case. If the experiment is  at $100$Hz, allowing a resolution of $10^{10}$, it detects the slow drift of lines due to cosmological redshifting by $\approx 10^{-10}$ each year or so, and possibly even resolves the intrinsic line broadening from quantum gravity. 

Given the current variety and state of change in models,  these estimates suggest that holographic discreteness from inflation should for the time being be considered part of the phenomenological repertoire of quantum gravity.

\begin{acknowledgements}
I am grateful to A. Karch for useful discussions and  M. Jackson for useful comments, and to  Fermilab for hospitality.
This work was supported by NSF grant AST-0098557 at the University of
Washington.

\end{acknowledgements}

\end{document}